\documentclass[a4paper,11pt]{article}

\usepackage[margin=1.2in]{geometry}
\usepackage{jheppub} 
\usepackage{graphicx}  
\usepackage{dcolumn}   
\usepackage{bm}        
\usepackage{amssymb}   
\usepackage{subcaption}
\usepackage{setspace}
\usepackage{amsmath}
\usepackage{amssymb}
\usepackage{fancybox}
\usepackage{xcolor}
\usepackage[toc,page]{appendix}

\usepackage[protrusion=false,tracking=false,spacing=false,factor=1000,stretch=10,shrink=40]{microtype}
\usepackage[english]{babel}
\usepackage{amsmath,amsthm,amssymb}
\usepackage{amsfonts}
\usepackage{amsopn}
\usepackage{float}
\usepackage{setspace}
\usepackage{fancybox}
\usepackage{url}
\usepackage{appendix}
\usepackage{graphicx}
\usepackage{dcolumn}
\usepackage{bm}
\usepackage{color}
\usepackage{esint}
\usepackage{tikz-feynman}
\usepackage{slashed}
\usepackage{booktabs}

\usepackage{float}
\usepackage{titlesec}

\newcommand{\ra}[1]{\renewcommand{\arraystretch}{#1}}
\newcommand{\beq}{\begin{equation}}
\newcommand{\eeq}{\end{equation}}
\newcommand{\ba}{\begin{array}}
\newcommand{\ea}{\end{array}}
\newcommand{\bea}{\begin{eqnarray}}
\newcommand{\eea}{\end{eqnarray} }
\newcommand{\be}{\begin{eqnarray}}
\newcommand{\ee}{\end{eqnarray} }

\newcommand{\bal}{\begin{align}}
\newcommand{\eal}{\end{align}}
\newcommand{\bi}{\begin{itemize}}
\newcommand{\ei}{\end{itemize}}
\newcommand{\ben}{\begin{enumerate}}
\newcommand{\een}{\end{enumerate}}
\newcommand{\bc}{\begin{center}}
\newcommand{\ec}{\end{center}}
\newcommand{\bt}{\begin{table}}
\newcommand{\et}{\end{table}}
\newcommand{\btb}{\begin{tabular}}
\newcommand{\etb}{\end{tabular}}

\newcommand{\ncomment}[1]{}

\numberwithin{equation}{section}
\title{Detecting Dark Photons from Atomic Rearrangement in the Galaxy}
\author{James Eiger, Michael Geller}
\affiliation{Department of Physics, Tel Aviv University, Tel Aviv, Israel} 

\abstract{We study a new dark sector signature for an atomic process of ``rearrangement" in the galaxy. In this process, a hydrogen-like atomic dark matter state together with its anti-particle can rearrange to form a highly-excited bound state. This bound state will then de-excite into the ground state emitting a large number of dark photons that can be measured in experiments on Earth through their kinetic mixing with the photon. We find that for DM masses in the GeV range, the dark photons have enough energy to pass the thresholds of neutrino observatories such as Borexino and Super-Kamiokande that can probe for our scenario even when our atomic states constitute a small fraction of the total DM abundance.  We study the corresponding bounds on the parameters of our model from current data as well as the prospects for future detectors.}

\begin{document}
\maketitle

\newpage
\section{Introduction}
  
Dark matter (DM) remains one of the greatest unsolved mysteries in modern physics. There is considerable evidence for its existence over many astrophysical scales, however, only very little is known about it beyond its gravitational interaction (see e.g. \cite{Bertone:2004pz,Jungman:1995df}). Various dark matter candidates have been proposed in literature (reviewed in \cite{Feng:2010gw}) and an experimental endeavor is underway to search for them. Most of the theoretical and experimental effort has been focused on the weakly interacting massive particle (WIMP) scenario where DM is heavy and can be searched for via nuclear recoil with heavy elements \cite{Akerib:2016vxi,Aprile:2018dbl}. In recent years the scope has been broadened from a single dark matter particle to a ``dark sector" which could encompass several particles and their interactions with themselves and with baryons (reviewed in \cite{Essig:2013lka}),  and the mass range has been expanded to include also light dark matter which is the focus of an extensive theoretical and experimental effort (see \cite{Knapen:2017xzo} for a review).

In this paper we consider a dark sector scenario with a unique phenomenology that results in an unusual experimental signature. In particular, we examine an atomic dark sector, where the constituents are a heavy $X$ and a light $\ell$ fermions together with their antiparticles, all charged under a dark $U(1)_{\rm bind}$. We will assume that the atomic states $X\bar{\ell}$ and $\bar{X}\ell$ make up some fraction of DM in galaxies. The unique phenomenology of these states is a process called ``rearrangement" that occurs when an $X\bar \ell$ and a $\bar{X} \ell$ collide\footnote{This process occurs similarly for an atomic state and a free corresponding $X$ or $\bar{X}$. }. This is similar to a SM process that occurs e.g. for Hydrogen and anti-Hydrogen\cite{Fermi:1947uv,PhysRev.77.521,PhysRevA.64.052712,PhysRevLett.84.4577,PhysRevA.11.1792,PhysRevLett.28.1227,PhysRevA.66.032506}: at a critical radius between the two $X$ nuclei, the binding energy of the two light $\ell$'s vanishes, resulting in their detachment from their respective heavy nucleus \cite{PhysRev.77.521}. The $X$ and $\bar{X}$ particles will then instantly join into a new, excited bound state. As these excited states de-excite, they emit highly energetic dark photons and upon reaching an angular momentum of zero, annihilate, producing a final dark photon pair with considerably higher energies.
 
As a portal to the dark sector, we will assume that the $X$s are additionally charged under a separate $U(1)_{\rm mix}$ that kinetically mixes with the SM photon \cite{Holdom:1985ag}. This choice, compared with coupling to the $U(1)_{\rm bind}$, has a larger unconstrained parameter space. With this portal, the dark photons produced in the rearrangement process can be observed in the large volume detectors of neutrino observatories\footnote{For other ideas of detecting DM with neutrino detectors see \cite{Eby:2019mgs,Geller:2020tyv,Grossman:2017qzw}.} through Compton scattering with the electrons in the detector and their subsequent recoil  \cite{Abe:2016nxk,Alimonti:2008gc,Anderson:2018ukb}. This will allow us to set an upper bound on the kinetic mixing parameter from the existing data, mainly from Borexino \cite{Collaboration:2011nga} and Super-Kamiokande \cite{2003NIMPA.501..418F}, and to project discovery bounds for future experiments - such as Hyper-Kamiokande \cite{hyperkamiok2018hyperkamiokande} and JUNO \cite{An:2015jdp}. We will compare these bounds to those from standard direct detection via the nuclear/electron recoil by the atomic states and to Supernova 1987A cooling constraints \cite{Chang:2018rso}.

This paper is organized as follows: in section~\ref{sec:model} we introduce our model; in section~\ref{sec:dynamics} we explore the dynamics of the atomic DM in this model; in section~\ref{sec:constraints} we study the constraints on our scenario and in section~\ref{sec:conclusion} we conclude.

\section{Model} \label{sec:model}
\begin{table}
\ra{1.3}
\begin{center}
 \begin{tabular}{@{}lll@{}} 
 \toprule
 Particle & $U(1)_{\rm mix}$ & $U(1)_{\rm bind}$ \\ [0.5ex] 
 \toprule
 $X\bar{X}$ & 1, -1 & 1, -1 \\
 \hline
 $\ell\bar{\ell}$ & 0 & 1, -1 \\
 \bottomrule
\end{tabular}
\end{center}
  \caption{The field content of the dark sector with the $U(1)$ charges at tree level.}
  \label{Tab1}
\end{table}

We consider a dark sector model with a heavy dark fermion $X$ and a light dark fermion $\ell$, constituting a fraction $f_X$ of the total DM. These fermions are charged under two $U(1)$  gauge groups: $U(1)_{\rm bind}$ which allows $\bar{X}\ell$ and $X\bar \ell$ atoms to form and $U(1)_{\rm mix}$ which is kinetically mixed with the SM and serves as a portal to the dark sector.  An overview of this can be seen in Table~\ref{Tab1}. We will chose for simplicity that the couplings of both $U(1)$s are $\alpha_{\rm mix}=\alpha_{\rm bind}\equiv\alpha_D$. The Lagrangian of this dark sector is given by
\renewcommand{\arraystretch}{1.5}
 
\begin{eqnarray}
\mathcal{L}_{DS}=-\frac{1}{4}F_{\mu\nu}^{(i)}F^{\mu\nu}_{(i)}+\bar{X}(i\slashed{D}^{(i)}-m_X)X+\bar{\ell}(i\slashed{D}^{\rm bind}-m_l)\ell +\hspace{1mm}m_{\gamma, (i)}^2A_\mu^{(i)}A^\mu_{(i)}+\frac{\epsilon}{2} F_{\mu\nu}F^{\mu\nu}_{\rm mix}\,,\nonumber\\
\end{eqnarray}
where $i$ runs over the two gauge groups. For the masses of the dark photons we assume that $A_{\rm mix}$ has a small mass above inverse galactic scales and hence is not decoupled from the SM in its galactic propagation. The specific value of this mass does not enter in our calculations and we only assume that the mass is below $O(\rm eV)$ where the stringent limits disappear (see e.g. \cite{Barkana:2018cct}). We remain agnostic regarding the mass of $A_{\rm bind}$ as long as it is small enough to allow atoms to form.
  
In the mass basis, the kinetic mixing results in an interaction term of form
\beq
\mathcal{L}_{int}=\epsilon\bar{\psi}(i\slashed{D}^{\rm mix}-m)\psi,
\eeq
for any SM fermion $\psi$, and in particular, for an electron. This results in a $\epsilon^2$ suppressed scattering of the dark photon $A_{\rm mix}$/dark fermions ($\ell,X$) with the electrons in the detector (Klein-Nishina/Rutherford). 

\section{Dynamics of Dark Matter Bound States}\label{sec:dynamics}
Having outlined the model, we now move to discuss its phenomenology. We leave the discussion of early cosmology for a subsequent work and assume a symmetric or mostly symmetric population of $X$ and $\ell$ and their anti particles which altogether constitute some very small fraction $f_X$ of total dark matter.\footnote{The parameter space of interest in this work may require a non-standard production mechanism, e.g. via decays of a long lived particle. The early universe recombination dynamics in this dark sector are non-trivial and vary significantly from the SM due to the population of both $X$ and $\bar{X}$, and require an in-depth analysis which is left for a future work.}

\subsection{Rearrangement}\label{sbsec:rearrangement}
  We first review the rearrangement process at the heart of our dark sector phenomenology. For simplicity, we assume that all the dark $X$ and $\ell$  are in atomic $X\bar{\ell}$ and $\bar{X}l$ states and it is easy to extrapolate the results to the case of fractional atomic abundance. These atoms, upon meeting their anti particles, can "rearrange" into  $\bar{X}{X}$ and $\bar{\ell\ell}$ \cite{PhysRev.77.521} bound states. 

 The SM analog of this process has been widely studied in literature dating back to Fermi and Teller \cite{Fermi:1947uv} where a similar process of muon capture on Hydrogen, i.e. $\mu + H  \to H_{\mu} + e$, was analyzed using the adiabatic approach. Subsequently, the Hydrogen and anti-Hydrogen rearrangement, $H+\bar{H}\to p\bar{p} + e^+e^-$,  was calculated in     \cite{PhysRevA.64.052712,PhysRevLett.84.4577,PhysRevA.11.1792,PhysRevLett.28.1227,PhysRev.77.521,PhysRevA.66.032506}. Here we briefly review this analysis, adapting it to our scenario. In the adiabatic approximation the $X$s are assumed to be non dynamic and the energy is calculated by solving the Schrödinger equation for the $\ell$s while varying the inter-$X$ separation. As the  $X\bar{\ell}$ and $\bar{X}\ell$  approach one another, the binding energy of $\ell$ decreases until it vanishes at a critical distance $R_c$ 

\beq\label{Rcrit}
R_c\approx0.7a_0^{X\ell}=\frac{0.7}{\alpha_Dm_\ell}\,,
\eeq
where $a_0^{X\ell}$ is the Bohr radius of the $\bar{X}\ell$ atoms. Once the incoming $X$s come within this distance, the $\ell$s become no longer bound and they escape, usually in the form of the $\bar \ell \ell$ bound state. The remaining $X$ and $\bar{X}$ have a resulting excess of negative energy - and therefore become bound in an $\bar{X}{X}$ state. This state is created with the typical distance between the $X$ and $\bar{X}$ of $R_c$, which, in the case of a mass hierarchy, is much larger than the Bohr radius of the $X\bar{X}$ system given by:
\beq\label{bohrxx}
a_0^{X\bar{X}}=\frac{2}{\alpha_Dm_X}.
\eeq
Hence, in this case, the $X\bar{X}$ is produced in a highly excited state with the principle quantum number
\beq\label{nini}
n_{\rm ini}\approx\sqrt{\frac{m_X}{2m_\ell}}.
\eeq
This highly excited $\bar{X}{X}$ will subsequently de-excite, releasing dark photons in the process as the bound state makes the long fall to its ground state. While the annihilation between the $X$ and $\bar{X}$ within the bound state is possible, it is highly suppressed compared with the de-excitation rate at large angular momentum. We will take the simplistic assumption that each ($l,m$) state is produced at equal measure, which is not far from what is shown in the simulations (see e.g. \cite{PhysRevA.66.032506}). With this assumption, the high-$l$s dominate in this production process, as the number of $m$ states for each $l$ is $2l+1$.

To calculate the rearrangement rate we need to find the impact parameter such that the minimal distance between the $X$ and the $\bar{X}$ in their classical trajectory is $R_c$. Since both $X$ particles are charged under $U(1)_{\rm mix}$, there is a ``focusing" effect.  Using energy and angular momentum conservation for an initial velocity of $\langle v\rangle$, we find the total cross section for rearrangement is:

\beq\label{eq:rearrangement_xsec}
\sigma_{\rm rearr}=\pi b^2=\pi R_c^2\left(1+\frac{\alpha_{mix}}{R_cm_X\langle v\rangle^2}\right)
\eeq
 Correspondingly, the total production rate of $X\bar{X}$ bound states is
\beq\label{Z}
R_{\rm prod} =\int{\sigma_{\rm rearr} \sqrt{2} \langle v\rangle\left(\frac{f_X\cdot\rho(\vec{r})}{2m_X}\right)^2 d^3r}
\eeq

\subsection{Emission}

Having calculated the rearrangement cross section, from which we can infer the rate of $\bar{X}{X}$ production,  we move to calculate the rate of photon emission in the de-excitation of this state. The de-excitation rates can be calculated using Fermi's golden rule where we only consider electric dipole transitions, while higher multipole transitions are suppressed by powers of $\alpha_D$.
 
To calculate all the emission rates, we need to find the occupation numbers of any given state $\psi_{n,l,m}$ in the $\bar{X}{X}$ tower. We derive these numbers recursively, first considering the states at the initial energy level $n_{ini}$ where we assume that any given state of $l,m$ has an equal chance of being created. The production rate of any initial state is therefore 
$\frac{R_{\rm prod}}{\left(n_{\rm ini}+1\right)^2}$, where $R_{\rm prod}$ is the total production rate. We have checked that the collisional de-excitation rate is very low in the galaxy. The occupation numbers at the initial state then quickly reach the equilibrium values of
\beq\label{Nmaxsteadysol}
N_{n_{ini},l_1,m_1}^{\rm eq}=\frac{R_{\rm prod}}{\left(n_{\rm ini}+1\right)^2 \Gamma(l_1,m_1\to \rm all)}.
\eeq
For all the rest of the states, we can now calculate all the occupation numbers  by considering all the transitions to and from each ($n,l,m$) state: 
\beq
\sum_{n',l',m'}N_{n',l',m'}\Gamma\left( n',l',m' \to n,l,m \right)-N_{n,l,m} \Gamma \left(n,l,m \to \rm all \right)=0
\eeq
With this we can calculate the emission spectrum by calculating the rate of all the transitions multiplied by the occupation number. The emitted photons have a discrete spectrum based on the energy levels and transitions permitted by the selection rules, where the emission lines depend on $m_X$ and $\alpha_D$, while $n_{\rm ini}$ is determined by the ratio of the two masses. In this analysis we ignore the dark photons from annihilation. This allows us to remain open to models with more than a single type of $X$ particle, and furthermore, the number of dark photons generated by the annihilation is negligible compared to the dark photons from the de-excitation process.

\begin{figure*}[!ht]
    \includegraphics[width=1\linewidth]{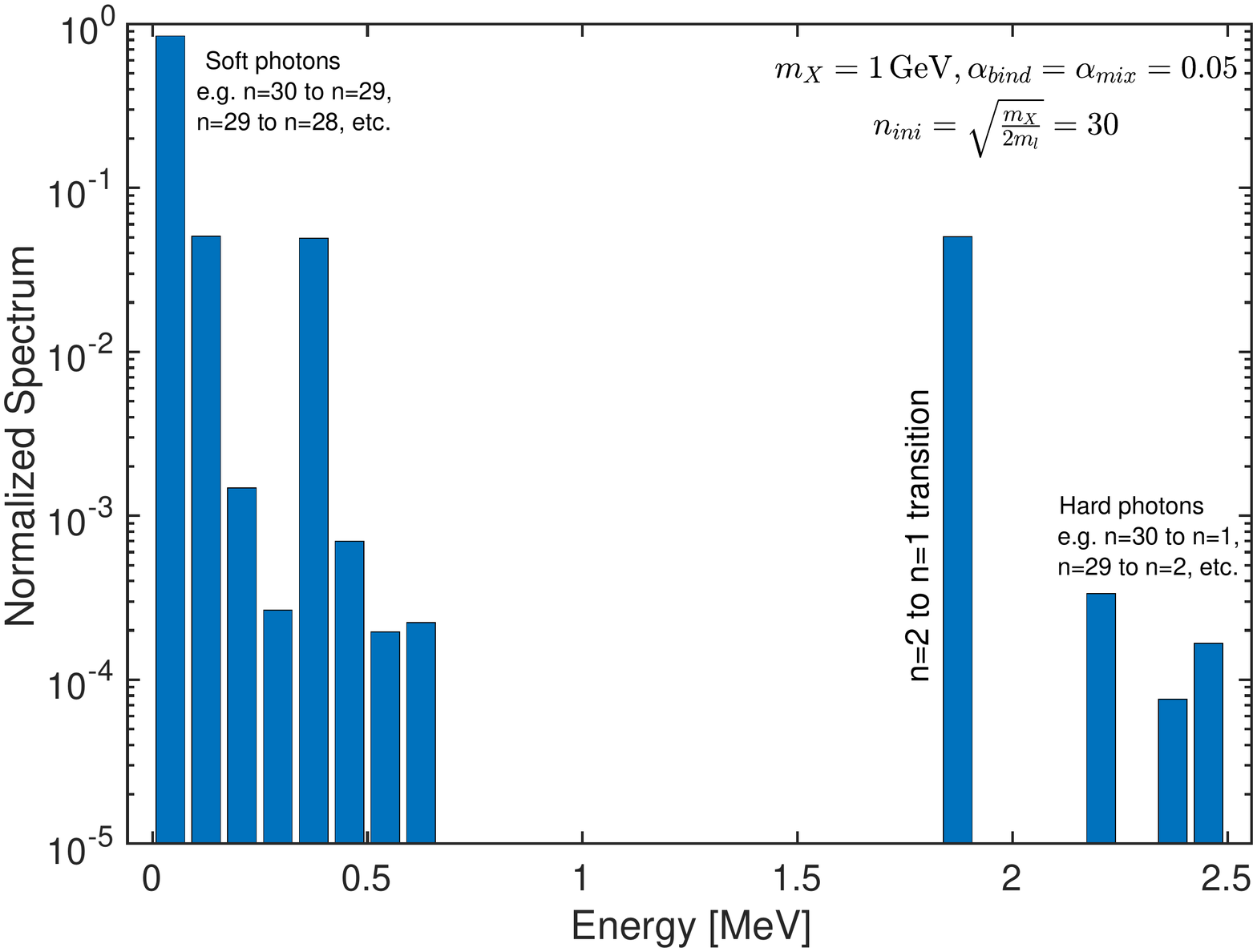}
  \caption{Example of a spectrum of emitted photons from the de-excitation of $X\bar{X}$, produced at an energy level of $n_{ini}=30$. Note that this does not include potential photons which could be generated from annihilation of both the $X\bar{X}$ and $l\bar{l}$.} 
  \label{spectrumcompare}
\end{figure*}
 We plot the spectrum for one choice of parameters in Fig.~\ref{spectrumcompare}. Here, the leftmost peaks correspond to the the soft dark photons in $n\rightarrow n-1$ transitions close to the initial state $n_{\rm ini}$. The most energetic peaks are due to direct de-excitations from $n_{\rm ini}$ to the lowest levels, but the "allowed" transition only exists for low angular momentum at the initial state and therefore they correspond to only a small fraction of the emitted dark photons. A notable energetic peak is due to the decay from $n=2$ into $n=1$, which is the last part of the decay chain for most of the initial states. We see that for our choice of $m_X=1$ GeV, a large fraction of the emitted dark photons are highly energetic and pass the thresholds for detection at neutrino experiments.

\section{Constraints} \label{sec:constraints}
\begin{figure}[!ht]
\centering
\includegraphics[width=\linewidth]{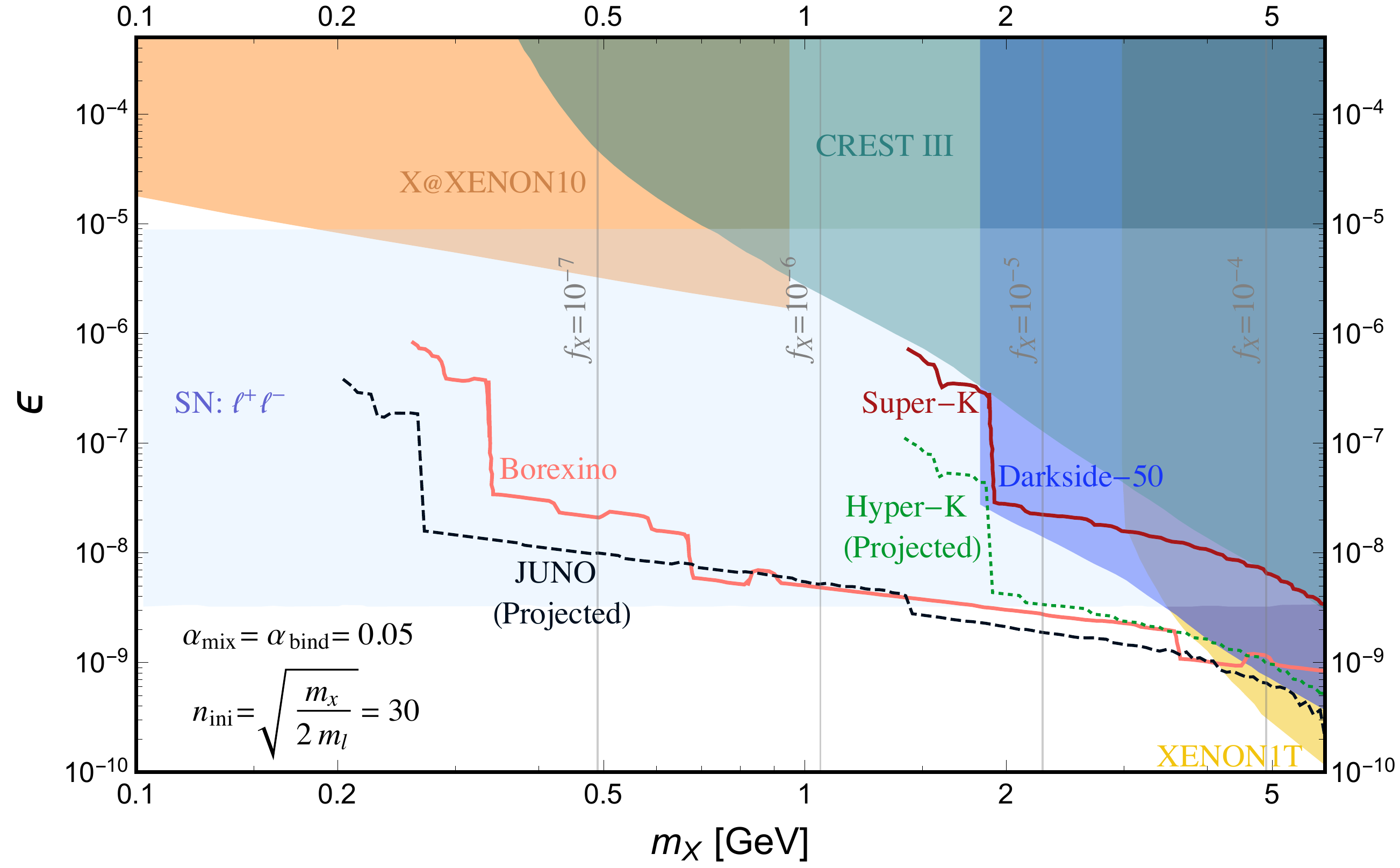}
\caption{Bounds on the kinetic mixing in our scenario from Super-Kamiokande  \cite{2003NIMPA.501..418F} and Borexino \cite{Collaboration:2011nga} experiments together with projections for Hyper-Kamiokande \cite{hyperkamiok2018hyperkamiokande} and Juno \cite{An:2015jdp}. Also plotted are bounds from direct detection of the atomic states from XENON10 \cite{Essig:2015cda}, CRESST \cite{Petricca:2017zdp}, Darkside \cite{Agnes:2018ves}, XENON1T \cite{Aprile:2019xxb} and finally the bounds from SN1987a \cite{Chang:2018rso}, which have been recently questioned \cite{Bar:2019ifz}. The fraction $f_X$ is chosen to be the limiting one beyond which our atomic states are depleted (given in the horizontal lines).} 
\label{all bounds}
\end{figure}

We can now calculate the bounds on our scenario, which will be given in terms of the kinetic mixing parameter $\epsilon$. We first focus on the new signature of dark photons from the de-excitation of $X\bar{X}$ which is produced via atomic rearrangement. These dark photons interact via Compton-like scattering with electrons and therefore can be searched for using electron-recoil in large volume neutrino detectors. For standard DM candidates which pass the detector at non-relativistic speeds, neutrino detectors cannot be used for direct detection due to their high threshold. In our case, however, the dark photons are highly relativistic and therefore can accelerate the electrons in the detector to above threshold energies. Furthermore, electron recoil is the primary signal in these detectors and so their results can be trivially adapted for our scenario. The relevant detectors are the Super-Kamiokande \cite{2003NIMPA.501..418F} and Borexino \cite{Alimonti:2008gc} experiments. Borexino, while being the smaller one, has a lower energy threshold and hence is effective in setting bounds for lower range of $m_X$. Super-Kamiokande is the larger, and is consequently expected to generate the most events at high masses, where we are already excluded by nuclear recoil experiments. We note that IceCube in particular (even DeepCore) \cite{Abbasi:2008aa}, while being the largest, is not compatible with this analysis due to its extremely high energy threshold. We plot the calculated bounds on the kinetic mixing $\epsilon$ from Super-Kamiokande and Borexino in Fig.~\ref{all bounds}.  We also plot the projections for future neutrino detectors, specifically JUNO \cite{An:2015jdp}  and Hyper-Kamiokande  \cite{hyperkamiok2018hyperkamiokande}. In these calculations the dark atomic fraction $f_X$ is chosen to be the highest possible fraction of $f_X(m_X)$ that will keep the rearrangement process from depleting all the $X\ell$ atoms, assuming there is no mechanism that replenishes them (the horizontal lines in Fig.~\ref{all bounds}). To extrapolate to any other fraction, we remind the reader that the number of emitted dark photons scales as $f^2_X$ and therefore the bounds and projections on $\epsilon$ from the neutrino experiments scale linearly with $f_X$.

The bounds are calculated assuming a NFW dark matter density profile \cite{Navarro:1995iw} which gives the most stringent bounds due to its high density in the core. However, the difference with a cored distribution is marginal, due to the core being a relatively small region of the total dark matter halo. To illustrate this we compare the average DM density squared in a cuspy NFW profile and a cored Burkert profile \cite{Burkert:1995yz,Salucci:2000ps}, and the ratio of the two turns out to be merely $\frac{\langle n_{\rm NFW}^2\rangle}{\langle n_{\rm Burkert}^2\rangle}=1.72$. The ratio of the $X$ and $\ell$ masses is chosen to be the same as the proton to electron mass ratio and the fine structure constant of the two $U(1)$s is taken to be $\alpha_{\rm mix}=\alpha_{\rm bind}=0.05$.

More standard bounds come from the direct detection of our atomic states $X\bar{\ell}$, mainly via electron and nuclear recoil in the mass range of interest (see \cite{Essig:2011nj,Essig:2015cda,Petricca:2017zdp,Agnes:2018ves,Aprile:2019xxb}). Note that the since the atomic state is milli-charged, they scatter with approximately the same cross-section and form factors as free $X$. We also consider the SN1987A bounds due to the production and subsequent escape of the light $\ell$ states \cite{Chang:2018rso} which couple to the SM through the loop-induced mixing of the $U(1)_{\rm bind}$. 

In this scenario, the SN bounds are competitive with the neutrino detector bounds, but it is worthwhile to note that these constraints may well be dependent on the mechanism of the supernova explosion, as was explored in \cite{Bar:2019ifz} and thus it is important to consider terrestrial experiments. Future neutrino detectors may go further beyond the SN bounds. As for neutrino experiments, these bounds are calculated for the maximal fraction  $f_X(m_X)$, and to extrapolate to other values, we note that the bounds from direct detection of the atomic states scale as $\sqrt{f_X}$ and the SN bounds are independent of the dark atomic fraction.

\section{Conclusions}\label{sec:conclusion}
In this paper we have presented a new dark sector phenomenology, based the process of atomic rearrangement. In this process, hydrogen-like atomic states can re-arrange with their anti particles to form a highly excited protonium-like bound state. The subsequent de-excitation process of this bound state releases high-energy dark photons which can then be observed in experiments via their kinetic mixing with the SM photon. We have shown that neutrino detectors are sensitive to these signals and calculated the bounds using existing data from Super-Kamiokande and Borexino. Future experiments, such as Hyper-Kamiokande and JUNO would be able to probe further regions of the parameter space, and in the case of discovery could confirm this scenario via the smoking gun of the discrete spectrum determined by the atomic lines of the excited state. 

\section{Acknowledgment} 

We thank Ofri Telem for comments on the draft. We also acknowledge useful comments from Itay M. Bloch, Shmuel Nussinov, Nadav Outmezguine and Tomer Volansky.
M.G. and J.E are supported by the Israel Science Foundation (Grant No. 1302/19) and by the US-Israeli BSF (grant 2018236). MG thanks KITP at Santa Barbara and MIAPP for hospitality while this work was in progress.

\bibliography{ref}

\providecommand{\href}[2]{#2}\begingroup\raggedright\begin{thebibliography}{10}

\bibitem{Bertone:2004pz}
G.~Bertone, D.~Hooper, and J.~Silk, {\it {Particle dark matter: Evidence,
  candidates and constraints}},  {\em Phys. Rept.} {\bf 405} (2005) 279--390,
  [\href{http://arxiv.org/abs/hep-ph/0404175}{{\tt hep-ph/0404175}}].

\bibitem{Jungman:1995df}
G.~Jungman, M.~Kamionkowski, and K.~Griest, {\it {Supersymmetric dark matter}},
   {\em Phys. Rept.} {\bf 267} (1996) 195--373,
  [\href{http://arxiv.org/abs/hep-ph/9506380}{{\tt hep-ph/9506380}}].

\bibitem{Feng:2010gw}
J.~L. Feng, {\it {Dark Matter Candidates from Particle Physics and Methods of
  Detection}},  {\em Ann. Rev. Astron. Astrophys.} {\bf 48} (2010) 495--545,
  [\href{http://arxiv.org/abs/1003.0904}{{\tt arXiv:1003.0904}}].

\bibitem{Akerib:2016vxi}
{\bf LUX} Collaboration, D.~S. Akerib et~al., {\it {Results from a search for
  dark matter in the complete LUX exposure}},  {\em Phys. Rev. Lett.} {\bf 118}
  (2017), no.~2 021303, [\href{http://arxiv.org/abs/1608.07648}{{\tt
  arXiv:1608.07648}}].

\bibitem{Aprile:2018dbl}
{\bf XENON} Collaboration, E.~Aprile et~al., {\it {Dark Matter Search Results
  from a One Ton-Year Exposure of XENON1T}},  {\em Phys. Rev. Lett.} {\bf 121}
  (2018), no.~11 111302, [\href{http://arxiv.org/abs/1805.12562}{{\tt
  arXiv:1805.12562}}].

\bibitem{Essig:2013lka}
R.~Essig et~al., {\it {Working Group Report: New Light Weakly Coupled
  Particles}},  \href{http://arxiv.org/abs/1311.0029}{{\tt arXiv:1311.0029}}.

\bibitem{Knapen:2017xzo}
S.~Knapen, T.~Lin, and K.~M. Zurek, {\it {Light Dark Matter: Models and
  Constraints}},  {\em Phys. Rev.} {\bf D96} (2017), no.~11 115021,
  [\href{http://arxiv.org/abs/1709.07882}{{\tt arXiv:1709.07882}}].

\bibitem{Fermi:1947uv}
E.~Fermi and E.~Teller, {\it {The capture of negative mesotrons in matter}},
  {\em Phys. Rev.} {\bf 72} (1947) 399--408.

\bibitem{PhysRev.77.521}
A.~S. Wightman, {\it Moderation of negative mesons in hydrogen i: Moderation
  from high energies to capture by an ${\mathrm{h}}_{2}$ molecule},  {\em Phys.
  Rev.} {\bf 77} (Feb, 1950) 521--528.

\bibitem{PhysRevA.64.052712}
S.~Jonsell, A.~Saenz, P.~Froelich, B.~Zygelman, and A.~Dalgarno, {\it Stability
  of hydrogen-antihydrogen mixtures at low energies},  {\em Phys. Rev. A} {\bf
  64} (Oct, 2001) 052712.

\bibitem{PhysRevLett.84.4577}
P.~Froelich, S.~Jonsell, A.~Saenz, B.~Zygelman, and A.~Dalgarno, {\it
  Hydrogen-antihydrogen collisions},  {\em Phys. Rev. Lett.} {\bf 84} (May,
  2000) 4577--4580.

\bibitem{PhysRevA.11.1792}
W.~Kolos, D.~L. Morgan, D.~M. Schrader, and L.~Wolniewicz, {\it
  Hydrogen-antihydrogen interactions},  {\em Phys. Rev. A} {\bf 11} (Jun, 1975)
  1792--1796.

\bibitem{PhysRevLett.28.1227}
B.~R. Junker and J.~N. Bardsley, {\it Hydrogen-antihydrogen interactions},
  {\em Phys. Rev. Lett.} {\bf 28} (May, 1972) 1227--1229.

\bibitem{PhysRevA.66.032506}
K.~Sakimoto, {\it Antiproton, kaon, and muon capture by atomic hydrogen},  {\em
  Phys. Rev. A} {\bf 66} (Sep, 2002) 032506.

\bibitem{Holdom:1985ag}
B.~Holdom, {\it {Two U(1)'s and Epsilon Charge Shifts}},  {\em Phys. Lett.}
  {\bf 166B} (1986) 196--198.

\bibitem{Eby:2019mgs}
J.~Eby, P.~J. Fox, R.~Harnik, and G.~D. Kribs, {\it {Luminous Signals of
  Inelastic Dark Matter in Large Detectors}},
  \href{http://arxiv.org/abs/1904.09994}{{\tt arXiv:1904.09994}}.

\bibitem{Geller:2020tyv}
M.~Geller and O.~Telem, {\it {Self Destructing Atomic DM}},
  \href{http://arxiv.org/abs/2001.11514}{{\tt arXiv:2001.11514}}.

\bibitem{Grossman:2017qzw}
Y.~Grossman, R.~Harnik, O.~Telem, and Y.~Zhang, {\it {Self-Destructing Dark
  Matter}},  {\em JHEP} {\bf 07} (2019) 017,
  [\href{http://arxiv.org/abs/1712.00455}{{\tt arXiv:1712.00455}}].

\bibitem{Abe:2016nxk}
{\bf Super-Kamiokande} Collaboration, K.~Abe et~al., {\it {Solar Neutrino
  Measurements in Super-Kamiokande-IV}},  {\em Phys. Rev.} {\bf D94} (2016),
  no.~5 052010, [\href{http://arxiv.org/abs/1606.07538}{{\tt
  arXiv:1606.07538}}].

\bibitem{Alimonti:2008gc}
{\bf Borexino} Collaboration, G.~Alimonti et~al., {\it {The Borexino detector
  at the Laboratori Nazionali del Gran Sasso}},  {\em Nucl. Instrum. Meth.}
  {\bf A600} (2009) 568--593, [\href{http://arxiv.org/abs/0806.2400}{{\tt
  arXiv:0806.2400}}].

\bibitem{Anderson:2018ukb}
{\bf SNO+} Collaboration, M.~Anderson et~al., {\it {Measurement of the $^8$B
  solar neutrino flux in SNO+ with very low backgrounds}},  {\em Phys. Rev.}
  {\bf D99} (2019), no.~1 012012, [\href{http://arxiv.org/abs/1812.03355}{{\tt
  arXiv:1812.03355}}].

\bibitem{Collaboration:2011nga}
{\bf Borexino} Collaboration, G.~Bellini et~al., {\it {First evidence of pep
  solar neutrinos by direct detection in Borexino}},  {\em Phys. Rev. Lett.}
  {\bf 108} (2012) 051302, [\href{http://arxiv.org/abs/1110.3230}{{\tt
  arXiv:1110.3230}}].

\bibitem{2003NIMPA.501..418F}
S.~Fukuda et~al., {\it {The Super-Kamiokande detector}},  {\em Nuclear
  Instruments and Methods in Physics Research A} {\bf 501} (Apr., 2003)
  418--462.

\bibitem{hyperkamiok2018hyperkamiokande}
H.-K. Proto-Collaboration et~al., {\it Hyper-kamiokande design report},  2018.

\bibitem{An:2015jdp}
{\bf JUNO} Collaboration, F.~An et~al., {\it {Neutrino Physics with JUNO}},
  {\em J. Phys. G} {\bf 43} (2016), no.~3 030401,
  [\href{http://arxiv.org/abs/1507.05613}{{\tt arXiv:1507.05613}}].

\bibitem{Chang:2018rso}
J.~H. Chang, R.~Essig, and S.~D. McDermott, {\it {Supernova 1987A Constraints
  on Sub-GeV Dark Sectors, Millicharged Particles, the QCD Axion, and an
  Axion-like Particle}},  {\em JHEP} {\bf 09} (2018) 051,
  [\href{http://arxiv.org/abs/1803.00993}{{\tt arXiv:1803.00993}}].

\bibitem{Barkana:2018cct}
R.~Barkana, N.~J. Outmezguine, D.~Redigolo, and T.~Volansky, {\it {Strong
  constraints on light dark matter interpretation of the EDGES signal}},  {\em
  Phys. Rev.} {\bf D98} (2018), no.~10 103005,
  [\href{http://arxiv.org/abs/1803.03091}{{\tt arXiv:1803.03091}}].

\bibitem{Abbasi:2008aa}
{\bf IceCube} Collaboration, R.~Abbasi et~al., {\it {The IceCube Data
  Acquisition System: Signal Capture, Digitization, and Timestamping}},  {\em
  Nucl. Instrum. Meth.} {\bf A601} (2009) 294--316,
  [\href{http://arxiv.org/abs/0810.4930}{{\tt arXiv:0810.4930}}].

\bibitem{Navarro:1995iw}
J.~F. Navarro, C.~S. Frenk, and S.~D.~M. White, {\it {The Structure of cold
  dark matter halos}},  {\em Astrophys. J.} {\bf 462} (1996) 563--575,
  [\href{http://arxiv.org/abs/astro-ph/9508025}{{\tt astro-ph/9508025}}].

\bibitem{Burkert:1995yz}
A.~Burkert, {\it {The Structure of dark matter halos in dwarf galaxies}},  {\em
  IAU Symp.} {\bf 171} (1996) 175,
  [\href{http://arxiv.org/abs/astro-ph/9504041}{{\tt astro-ph/9504041}}].

\bibitem{Salucci:2000ps}
P.~Salucci and A.~Burkert, {\it {Dark matter scaling relations}},  {\em
  Astrophys. J. Lett.} {\bf 537} (2000) L9--L12,
  [\href{http://arxiv.org/abs/astro-ph/0004397}{{\tt astro-ph/0004397}}].

\bibitem{Essig:2011nj}
R.~Essig, J.~Mardon, and T.~Volansky, {\it {Direct Detection of Sub-GeV Dark
  Matter}},  {\em Phys. Rev.} {\bf D85} (2012) 076007,
  [\href{http://arxiv.org/abs/1108.5383}{{\tt arXiv:1108.5383}}].

\bibitem{Essig:2015cda}
R.~Essig, M.~Fernandez-Serra, J.~Mardon, A.~Soto, T.~Volansky, and T.-T. Yu,
  {\it {Direct Detection of sub-GeV Dark Matter with Semiconductor Targets}},
  {\em JHEP} {\bf 05} (2016) 046, [\href{http://arxiv.org/abs/1509.01598}{{\tt
  arXiv:1509.01598}}].

\bibitem{Petricca:2017zdp}
{\bf CRESST} Collaboration, F.~Petricca et~al., {\it {First results on low-mass
  dark matter from the CRESST-III experiment}},  {\em J. Phys. Conf. Ser.} {\bf
  1342} (2020), no.~1 012076, [\href{http://arxiv.org/abs/1711.07692}{{\tt
  arXiv:1711.07692}}].

\bibitem{Agnes:2018ves}
{\bf DarkSide} Collaboration, P.~Agnes et~al., {\it {Low-Mass Dark Matter
  Search with the DarkSide-50 Experiment}},  {\em Phys. Rev. Lett.} {\bf 121}
  (2018), no.~8 081307, [\href{http://arxiv.org/abs/1802.06994}{{\tt
  arXiv:1802.06994}}].

\bibitem{Aprile:2019xxb}
{\bf XENON} Collaboration, E.~Aprile et~al., {\it {Light Dark Matter Search
  with Ionization Signals in XENON1T}},  {\em Phys. Rev. Lett.} {\bf 123}
  (2019), no.~25 251801, [\href{http://arxiv.org/abs/1907.11485}{{\tt
  arXiv:1907.11485}}].

\bibitem{Bar:2019ifz}
N.~Bar, K.~Blum, and G.~D'Amico, {\it {Is there a supernova bound on axions?}},
   {\em Phys. Rev. D} {\bf 101} (2020), no.~12 123025,
  [\href{http://arxiv.org/abs/1907.05020}{{\tt arXiv:1907.05020}}].

\end{thebibliography}\endgroup
\bibliographystyle{JHEP}
\end{document}